# ARTICLE

# Effect of Grain Coalescence on Dislocation and Stress Evolution of GaN Films Grown on Nanoscale Patterned Sapphire Substrates

Zuojian Pan,[a] Zhizhong Chen,[*abc] Yiyong Chen,[a] Haodong Zhang,[a] Han Yang,[a] Jingxin Nie,[a] Chuhan Deng,[a] Boyan Dong,[a] Daqi Wang,[a] Yuchen Li,[a] Weihua Chen,[a] Fei Jiao,[ad] , Xiangning Kang,[a] Chuanyu Jia,[e] Zhiwen Liang,[b] Qi Wang,[b] Guoyi Zhang[ab] and Bo Shen[ac]

Two types of nucleation layers (NLs), including in-situ low-temperature grown GaN (LT-GaN) and ex-situ sputtered physical vapor deposition AlN (PVD-AlN), are applied on cone-shaped nanoscale patterned sapphire substrate (NPSS). The initial growth process of GaN on these two NLs is comparably investigated by a series of growth interruptions. The coalescence process of GaN grains is modulated by adjusting the three-dimensional (3D) temperatures. The results indicate that higher 3D temperatures reduce the edge dislocation density while increasing the residual compressive stress in GaN films. Compared to the LT-GaN NLs, the PVD-AlN NLs effectively resist Ostwald ripening and facilitate the uniform growth of GaN grains on NPSS. Furthermore, GaN films grown on NPSS with PVD-AlN NLs exhibit a reduction of over 50% in both screw and edge dislocation densities compared to those grown on LT-GaN NLs. Additionally, PVD-AlN NLs result in an increase of about 0.5 GPa in the residual compressive stress observed in GaN films.

## Introduction

In response to the growing demand for high-definition display technology, light-emitting diodes (LEDs) are constantly developing in the direction of miniaturization[1-3]. For display technologies like virtual reality (VR) and augmented reality (AR), micro-LEDs with chip sizes smaller than 5 μm are typically necessary[4, 5]. However, the currently commercial microscale PSS (MPSS) exhibits a period of approximately 3 μm, which already approaches the chip size of micro-LEDs. MPSS can cause inhomogeneous light emission for ultra-small micro-LEDs, considering the rough interface after stripping the MPSS during the micro-LED chip process. Consequently, deploying nanoscale patterned sapphire substrates (NPSSs) has emerged as a crucial requirement for ultra-small micro-LED display technologies.

Defects and stress control of GaN films on NPSS are critical factors affecting the non-radiative recombination and polarization electric field of micro-LEDs[6, 7]. Importantly, the nucleation and coalescence process of GaN grains in the initial growth stages directly affects the crystal quality and stress distribution of GaN films on NPSS. Some studies have focused on the growth mechanism of GaN films on NPSS when using low-temperature grown GaN (LT-GaN) nucleation layers (NLs)[8, 9]. In recent years, physical vapor deposition AlN (PVD-AlN) has been developed to substitute LT-GaN as NLs to acquire high-quality GaN films[10-12]. High crystalline quality epitaxial growth of GaN films on MPSS with PVD-AlN NLs has been widely reported[12-18]. Past studies indicate a drastic reduction in dislocations at the top of the cone-shaped MPSS pattern when using PVD-AlN NLs compared to LT-GaN NLs[15, 16]. Nevertheless, few studies have been conducted on the growth of GaN films on NPSS using PVD-AlN NLs.

The coalescence process of GaN grains significantly influences the dislocation formation and stress evolution in GaN films on NPSS[9, 19-21]. During coalescence, tilting and twisting of GaN grains lead to screw and edge dislocations forming at the coalescence boundaries[21-23]. Meanwhile, grain coalescence introduces tensile stresses in GaN films[19, 20]. Notably, the density of GaN grains coalescence on NPSS at the top of the cone-shaped pattern increases by orders of magnitude compared to MPSS, which poses challenges in controlling dislocations at the coalescence boundaries[24, 25]. Nonetheless, recent research has revealed that precisely controlling grain orientation by nanopatterned substrates can almost eliminate dislocations in AlN films at the coalescence boundaries[26]. Accordingly, further investigations focusing on the coalescence of GaN grains on NPSS are required to realize high-quality GaN films.

In this paper, the coalescence process of GaN grains on NPSS with LT-GaN NLs and PVD-AlN NLs was studied in detail. Before grain coalescence, the grain size and density of the three-dimensional (3D) GaN layer were modulated by adjusting the 3D temperature. The surface morphology and orientation characteristics of the 3D GaN grains on NPSS with LT-GaN NLs and PVD-AlN NLs were comparably investigated. After completion of the GaN film growth, dislocation densities and stress states were evaluated for GaN films undergoing different grain coalescence processes. Furthermore, microscopic measurements were conducted to explore dislocations and stress distributions at the coalescence boundaries.

[a.] State Key Laboratory for Artificial Microstructure and Mesoscopic Physics, School of Physics, Peking University, Beijing 100871, China
E-mail: zzchen@pku.edu.cn
[b.] Dongguan Institute of Optoelectronics, Peking University, Dongguan, Guangdong 523808, China
[c.] Yangtze Delta Institute of Optoelectronics, Peking University, Nantong, Jiangsu 226000, China
[d.] State Key Laboratory of Nuclear Physics and Technology, School of Physics, Peking University, Beijing 100871, China
[e.] School of Electrical Engineering and Intelligentization, Dongguan University of Technology, Dongguan 523808, China

# ARTICLE

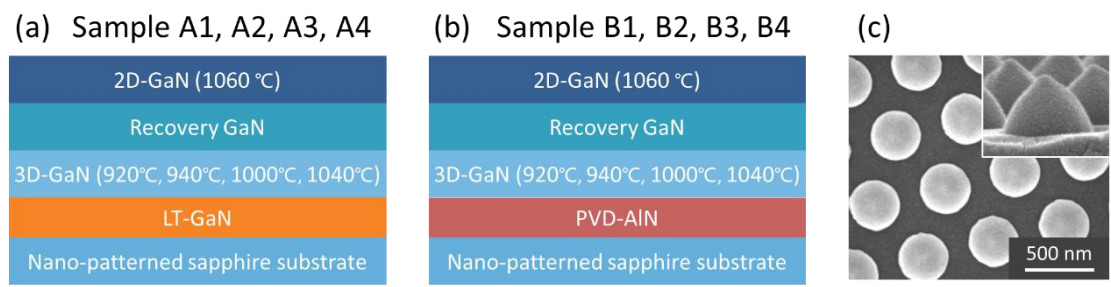

Fig. 1 Schematic diagrams of GaN films with (a) an in-situ LT-GaN NL (sample A1, A2, A3, A4) and (b) an ex-situ PVD-AlN NL (sample B1, B2, B3, B4). The order from 1 to 4 represents a gradual decrease in 3D temperature. (c) A plan-view SEM image of NPSS, and the inset shows a cross-sectional SEM image of the NPSS.

## Experimental

The 2-inch cone-shaped NPSS were fabricated by nano-imprint lithography (NIL) and inductively coupled plasma (ICP) dry etching using c-plane sapphire substrates. The diameter, spacing and height of the patterns were 400, 135 and 300 nm, respectively. The GaN films were grown on the prepared NPSS in an Aixtron Crius I 31*2 inch close-coupled showerhead metal organic chemical vapor deposition (MOCVD) system. Trimethylgallium (TMGa) and ammonia ($NH_3$) were used as group III and V precursors. Hydrogen ($H_2$) was used as the carrier gas. As shown in Fig. 1a-b, for sample A, a conventional in-situ 30 nm-thick LT-GaN NL was grown at 540°C on NPSS in MOCVD. For sample B, an ex-situ 30nm-thick PVD-AlN NL was deposited on NPSS using a NMC iTops A330 sputtering system. After that, 3D GaN layers were grown on both LT-GaN and PVD-AlN NLs with different 3D temperatures. Four sets of 3D temperatures (920 °C, 960 °C, 1000 °C, 1040 °C) were performed respectively for sample A1-A4 and sample B1-B4. Benefiting from the short coalescence process of GaN grains grown on NPSS, the 3D growth time of all the samples was set to 300s, which is sufficient for GaN grains to grow beyond cone height. Upon completion of the growth of the 3D GaN layer, GaN recovery layers were introduced to promote lateral coalescence. During growth of the recovery layers, the growth temperatures, pressures and V/III mole ratios transitioned slowly from the values of 3D process to those of three-dimensional (2D) process. Finally, the samples were finished with 2 μm-thick 2D GaN layers deposited at 1060 °C. The growth pressures of the 3D layer and 2D layer were 550 and 150 mbar. The V/III mole ratios of the 3D layer and 2D layer were 450 and 1022, respectively.

The features of 3D growth GaN islands were recorded using scanning electron microscopy (SEM) (FEI Navo NanoSEM 430). The nanoscale surface morphologies of the 3D GaN grains were further characterized using atomic force microscopy (AFM) (Bruker Dimension Icon). Ultra-fine AFM probes (Bruker FIB3D2-100A) with a high aspect ratio were utilized to obtain the exact morphology of the GaN grains. Besides this, the crystalline qualities of GaN coalesced films were measured using X-ray diffraction (XRD) (Panalytical X'Pert3 MRD) with a 1.5406 Å Cu Kα X-ray source operating at 40 kV and 40 mA. Transmission electron microscopy (TEM) (FEI Tecnai F20) was applied to study the mechanisms of dislocation formation and evolution. Then, the residual stress of the GaN epilayers was characterized by Raman scattering spectra (Horiba HR Evolution) with 0.5 cm$^{-1}$ spectral resolution. In addition, the stress distribution of GaN films was measured by Raman mapping (Witec Alpha300 R) with a spatial resolution of 300 nm.

## Results and discussion

### Initial growth process of GaN on NPSS

The distribution of 3D GaN islands could be affected by surface diffusion and desorption of adatoms. The growth temperatures, pressures and V/III mole ratios will affect the surface diffusion and desorption of Ga adatoms during 3D process[27, 28]. At relatively high temperatures, the surface diffusion and desorption rates of Ga adatoms are greatly increased[27, 28]. In this work, various 3D temperatures (920 °C, 960 °C, 1000 °C, 1040 °C) were applied to explore the coalescence process of GaN grown on NPSS. The distribution of 3D GaN islands on NPSS with LT-GaN and PVD-AlN NLs was analysed using SEM and AFM. From the SEM and AFM images, it can be clearly seen that 3D GaN islands arise from the c-plane trenches between the patterns and coalesces with the surrounding GaN islands, while the growth of GaN grains on the cone-shaped patterns is effectively suppressed.

ARTICLE

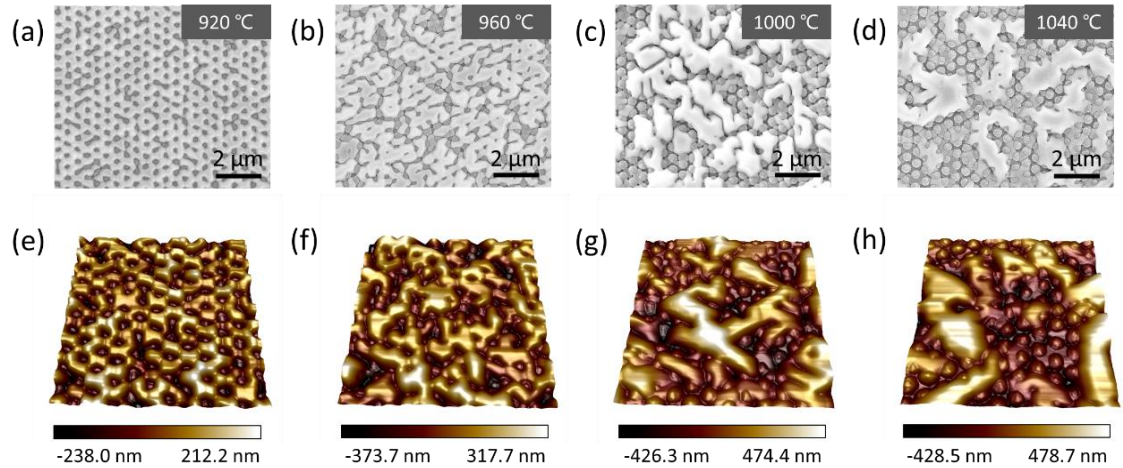

Fig. 2 Plan-view SEM images of interrupted samples after 3D GaN growth on NPSS using LT-GaN NLs, corresponding to 3D temperatures of (a) 920 °C, (b) 960 °C, (c) 1000 °C and (d) 1040 °C. All scale bars denote 2 μm. AFM images of these interrupted samples with 3D temperatures of (e) 920 °C, (f) 960 °C, (g) 1000 °C and (h) 1040 °C. The AFM images corresponds to 5 μm square regions.

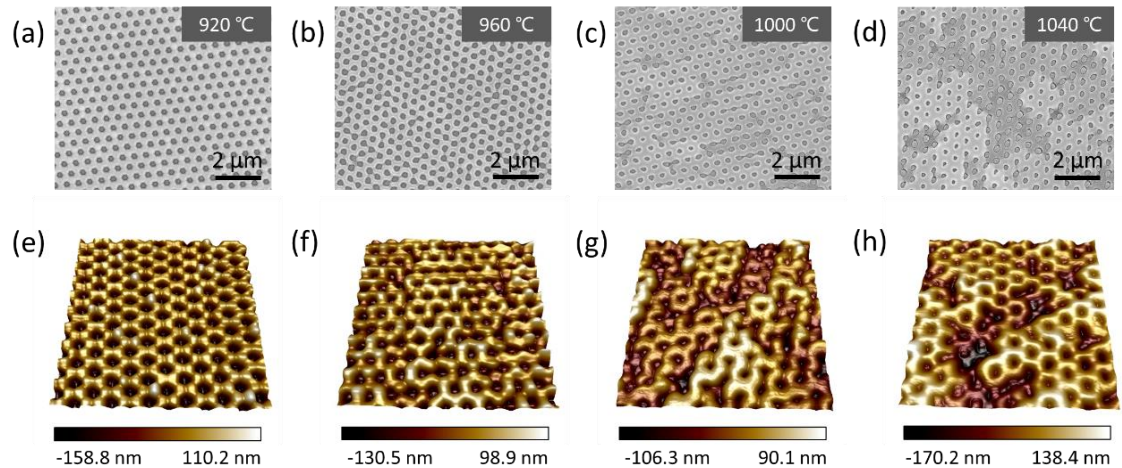

Fig. 3 Plan-view SEM images of interrupted samples after 3D GaN growth on NPSS using PVD-AlN NLs, corresponding to 3D temperatures of (a) 920 °C, (b) 960 °C, (c) 1000 °C and (d) 1040 °C. All scale bars denote 2 μm. AFM images of these interrupted samples with 3D temperatures of (e) 920 °C, (f) 960 °C, (g) 1000 °C and (h) 1040 °C. The AFM images corresponds to 5 μm square regions.

Fig. 2 shows the surface morphology of 3D GaN grains grown on NPSS with the LT-GaN NLs. The SEM and AFM images show that as the 3D temperatures increase from 920°C to 1040°C, the distribution of 3D GaN grains becomes inhomogeneous, accompanied by a reduction in grain density and an increase in grain size. From the statistics of AFM results, the average heights of the 3D GaN grains of the samples A1-A4 are 393, 528, 616 and 687 nm, respectively. The root-mean-square (RMS) roughness values of these samples are 63, 95, 120 and 126 nm, respectively. These results indicate that a relatively low 3D temperature is beneficial to the uniform growth of GaN on NPSS with LT-GaN NLs.

Higher 3D temperatures exacerbate the surface diffusion and desorption of Ga adatoms[27, 28]. In Fig. 2d and 2h, severe Ostwald ripening of 3D GaN grains grown on NPSS was observed when the 3D temperature was 1040 °C, which quite differs from the situation observed on MPSS[29, 30]. This observation indicates that there exists competition between GaN grains in the trenches between neighboring patterns during the growth process. The competition leads to the evaporation of the small GaN grains due to the higher equilibrium vapor pressure and the further growth of the large GaN grains to micrometer sizes. In other words, the diffusion length of Ga adatoms exceeds the patterning period of the NPSS at high 3D temperatures, resulting in the failure of the nanopatterns to limit the diffusion of Ga adatoms. Nevertheless, as the 3D temperatures gradually decrease, the surface diffusion and desorption of Ga atoms are weakened. The decrease in 3D temperatures results in a reduction in the diffusion length of Ga adatoms. Fig. 2a and 2d demonstrate that as the 3D temperature decreased to 920 °C, the nanopatterns effectively restricted the diffusion and desorption of Ga adatoms, thus resulting in uniform growth of GaN grains.

Fig. 3 displays the SEM and AFM images of 3D GaN grains grown on NPSS with the PVD-AlN NLs. Impressively, GaN grains can achieve uniform growth on NPSS with the PVD-AlN NLs even at 3D temperatures up to 1040 °C, which is quite different from the situation using LT-GaN NLs. As seen in Fig. 3d and 3h, only slight Ostwald ripening of GaN grains was observed at 3D temperature of 1040 °C. The AFM statistics show that the average heights of the 3D GaN grains of the samples B1-B4 are 252, 250, 215 and 288 nm, and the RMS roughness values of these samples are 37, 41, 36 and 47 nm, respectively. These results reveal that as the 3D temperature varies from 920 to 1040 °C, the growth of GaN grains on NPSS with AlN NLs remains uniform.

For the same 3D temperature conditions, the only difference between samples A1-A4 and samples B1-B4 is the type of NLs. Compared with the LT-GaN NLs, the PVD-AlN NLs can substantially improve the uniformity of GaN grains on NPSS. Obviously, the surface diffusion and desorption barriers of Al adatoms on the sapphire (0001) surface are larger than those of Ga adatoms due to the higher AlN bond energy (2.88 eV) than GaN bond energy (2.24 eV)[31]. The strong Al-N bonding makes the PVD-AlN NLs less susceptible to Oswald ripening at high temperatures. Further, the surface diffusion and desorption of Ga adatoms on the AlN layer are also suppressed, which facilitates the uniform distribution of GaN grains[32-34]. Accordingly, the comparison indicates that the PVD-AlN NLs are effective in attenuating the surface diffusion and desorption of Ga adatoms, thus improving the uniformity of GaN grains on NPSSs.

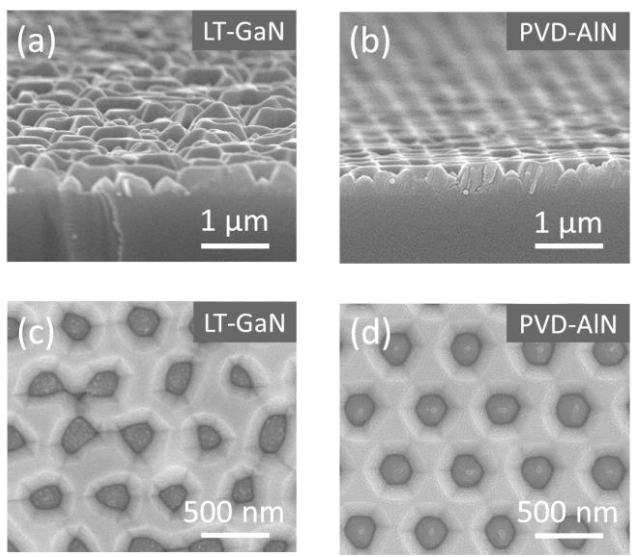

Fig. 4 Cross-section SEM images of the interrupted samples of sample A1 and B1 after 3D growth on NPSS with (a) LT-GaN and (b) PVD-AlN NLs. Plan-view SEM images of the interrupted samples of sample A1 and B1 after 3D growth on NPSS with (c) LT-GaN and (d) PVD-AlN NLs. The comparison of 4a and 4b reveal the disparity in the tilt (out-of-plane) of GaN grains, while the comparison of 4c and 4d demonstrate the disparity in the twist (in-plane) of GaN grains grown on NPSS with LT-GaN NLs and PVD-AlN NLs.

Grain coalescence process significantly affects dislocation formation and stress evolution in GaN films on sapphire substrates. The formation of screw and edge dislocations during grain coalescence is related to the tilting and twisting between GaN grain columns[22, 23]. Fig. 4 shows cross-sectional and plan-view SEM images of 3D growth-interrupted samples of samples A1 and B1 at a 3D temperature of 920°C. The tilt and twist of GaN grains correspond to the out-of-plane and in-plane misorientation. By comparing Fig. 4a and 4b, it is clear that GaN grains grown on NPSS with PVD-AlN NLs exhibit less tilt (out-of-plane) compared to those grown on LT-GaN NLs. Similarly, the comparison of Fig. 4c and 4d indicates that GaN grains on PVD-AlN NLs show less twist (in-plane).

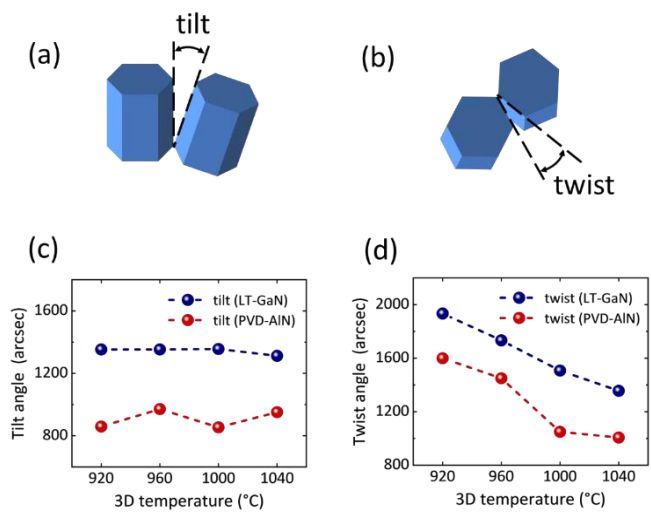

Fig. 5 Schematic diagram of (a) tilt angle and (b) twist angle between GaN grain columns. (c) Tilt and (d) twist angles of the samples A1-A4 and B1-B4 after 3D growth calculated from the ω-scans for symmetrical planes and ω/φ-scans for asymmetrical planes.

Fig. 5a-b illustrate the schematic diagram of tilt angle and twist angle between GaN grain columns. The average tilt and twist angles of GaN grain columns can be determined by XRD from the ω-scans at different symmetrical reflections and the ω/φ-scans at different asymmetrical reflections[35, 36]. The tilt angles $\beta_{tilt}$ between GaN grain columns can be acquired by ω-scans in the (00l) planes[37]. The dependence between the measured full-width at half-maximum (FWHM) of the ω-scan $\omega_{(00l)}$ and the tilt angles $\beta_{tilt}$ is shown below:

$$\omega^2_{(00l)} = \beta^2_{tilt} + k\left(\frac{1}{sin\theta_{(00l)}}\right)^2$$

where $\theta_{(00l)}$ is the measured peak position, and k is a coefficient related to the coherence lengths. One can see a linear relationship between $\omega^2_{(00l)}$ and $(1/sin\theta_{(00l)})^2$. Obviously we can use the intersection to derive the tilt angles.

Theoretically, the twist angles $\beta_{twist}$ between GaN grain columns can be acquired by ω/φ-scans perpendicular to the (00l) planes, with an inclination angle χ of 90 °. In fact, the twist angle can be accurately obtained by using ω-scans and φ-scans for planes corresponding to a χ approaching 90 °, as the (121) plane

 

corresponding to the χ of 78.6 °[38]. The twist angles can be taken as the average of FWHM of ω-scan and ϕ-scan from the (121) reflection:

$$\beta_{twist} = \frac{\omega_{(121)} + \varphi_{(121)}}{2}$$

Fig. 5c-d show the measured tilt and twist angles of 3D GaN grain columns grown on NPSS with the LT-GaN NLs and PVD-AlN NLs. As the 3D temperatures increase from 920°C to 1040°C, the tilt angles of GaN grains on NPSS have no significant change, while the twist angles gradually decrease. In fact, the spatial confinement effect of the periodic nanopatterns of NPSS is mainly in the dimension within the growth plane, whereas it is hardly to provide a confinement effect on the c-axis of the growth direction. In other words, the nanopatterns of NPSS are mainly effective in limiting the twisting of GaN grains, while hardly limiting the tilting of GaN grains. As a result, the tilt angle of the GaN grains remains almost unchanged at different 3D temperatures. From the SEM and AFM results, when the 3D temperatures increase from 920°C to 1040°C, the Ostwald ripening of GaN grains is observed, accompanied by a decrease in grain density and an increase in grain size, especially for the LT-GaN NLs. With smaller size and higher density, GaN grains are more prone to twist. When the size of the GaN grains increases, the GaN grains are less susceptible to twist due to their stronger connection to the substrate.

Comparison of LT-GaN and PVD-AlN NLs indicates that the PVD-AlN NLs drastically reduces the tilt and twist angles of GaN grains on NPSS. As shown in Fig. 5c-d, the average tilt angle of GaN grains grown on PVD-AlN NLs is reduced from 1345 to 910 arcsec, and the twist angles are also reduced by about 25% on average, compared to the LT-GaN NLs. These result are consistent with that observed in the SEM images of the GaN grains in Fig. 4. Actually, the LT-GaN NLs turns into small GaN grains after annealing at 3D temperature[9]. As a comparison, the PVD-ALN NLs exhibits better orientation uniformity and thermal stability compared to the LT-GaN NLs[15, 16]. Accordingly, the PVD-AlN NLs with high orientation uniformity can more effectively limit tilting and twisting of the 3D GaN grains.

**Dislocation and stress evolution of GaN films after coalescence**

Fig. 6 presents the in situ reflectance transients during the growth of GaN films for samples A4 and B4. From the SEM and AFM images in Fig. 2, GaN grains grown on the LT-GaN NLs suffer from severe Ostwald ripening. In this case, the GaN grains undergo an extended lateral growth process before coalescence occurs, which leads to a substantial increase in the coalescence thickness. The reflectance curve corresponding to the GaN layer grown on the LT-GaN NLs starts to oscillate after about 2000 s. In contrast, using PVD-AlN NLs significantly improves the growth uniformity of GaN grains, resulting in smaller GaN grain size and higher grain density, as seen in Fig. 3. The small-sized, high-density GaN grains rapidly coalesce after a short lateral growth process. The reflection curves of GaN layers grown on PVD-AlN NLs started oscillating after about 500 s. Therefore, the PVD-AlN NLs drastically reduce the coalescence thickness of GaN grains grown on NPSS compared to the LT-GaN NLs.

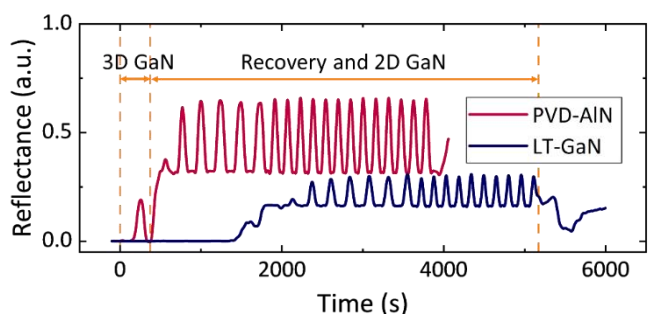

Fig. 6 In situ reflectance (at 405 nm) transients vs time during the growth of GaN films for samples A4 and B4.

To investigate the effect of the grain coalescence on dislocation formation in GaN films, XRD measurements were performed on samples A1-A4 and samples B1-B4. The edge and screw dislocation densities of each sample can be derived from the FWHM of the ω-scans in the (002) and (102) planes[36]. Fig. 7a-b demonstrate the edge and screw dislocation densities of GaN films on NPSS with LT-GaN and PVD-AlN NLs at different 3D temperatures. The results show that as the 3D temperatures increase, the screw dislocation density in GaN films remains essentially unchanged, while the edge dislocation density decreases significantly, as listed in Tab. 1. Combined with the surface morphology of 3D GaN grains before coalescence in Fig. 2-3, higher 3D growth temperatures lead to lower grain densities and larger grain sizes. Previous studies have shown that the coalescence of low-density and large-size GaN grains introduces less edge dislocation density due to smaller coalescence area[20, 39]. Moreover, combined with the tilt and twist angles of 3D GaN grains before coalescence, higher 3D temperatures can decrease the twisting of GaN grains but have no obvious restriction on GaN grain tilting. These observations imply that higher 3D temperatures facilitate the reduction of edge dislocation density, without affecting the screw dislocation density.

By comparing LT-GaN and PVD-AlN NLs, it is evident that the PVD-AlN NLs significantly reduce both the screw and edge dislocation densities in GaN films grown on NPSS. Specifically, the average screw dislocation density of GaN films grown on PVD-AlN NLs exhibited a reduction from $1.5*10^8$ cm$^{-2}$ to $5.0*10^7$ cm$^{-2}$, and the edge dislocation density is reduced by over 50% compared to LT-GaN NLs, as displayed in Fig. 7a-b. From the surface morphology of the 3D GaN grains in Fig. 3, the GaN grains grown on PVD-AlN NLs exhibit a high density of coalescence regions, mainly located in the top region of the cone-shaped pattern. Generally, a larger coalescence area corresponds to a higher density of dislocations[20, 24, 39]. Nevertheless, for GaN grains on PVD-AlN NLs, the high-density coalescence does not introduce excessive dislocations compared to LT-GaN NLs. Consequently, further exploration from a microscopic perspective is required to investigate the effect of the coalescence process on the dislocation formation.

ARTICLE

| | Sample | 3D temperature(°C) | XRD (002) | XRD (102) | Screw(cm$^{-2}$) | Edge(cm$^{-2}$) | Raman shift(cm$^{-1}$) | Residual Stress(Gpa) |
|---|---|---|---|---|---|---|---|---|
| NPSS/LT-GaN | A1 | 1040 | 256 | 361 | 1.3 E+8 | 1.1 E+9 | +2.5 | 0.98 |
| | A2 | 1000 | 253 | 400 | 1.3 E+8 | 1.4 E+9 | +2.2 | 0.86 |
| | A3 | 960 | 275 | 441 | 1.5 E+8 | 1.8 E+9 | +1.7 | 0.66 |
| | A4 | 920 | 283 | 536 | 1.7 E+8 | 2.8 E+9 | +1.4 | 0.55 |
| NPSS/PVD-AlN | B1 | 1040 | 149 | 225 | 4.5 E+7 | 4.4 E+8 | +3.6 | 1.41 |
| | B2 | 1000 | 171 | 258 | 5.9 E+7 | 6.4 E+8 | +3.4 | 1.33 |
| | B3 | 960 | 150 | 304 | 4.5 E+7 | 9.1 E+8 | +3.2 | 1.25 |
| | B4 | 920 | 159 | 342 | 5.1 E+7 | 1.2 E+9 | +2.8 | 1.09 |

Tab.1 Dislocation densities derived from XRD FWHMs and residual stresses obtained from Raman shifts of GaN films in the samples A1-A4 and B1-B4.

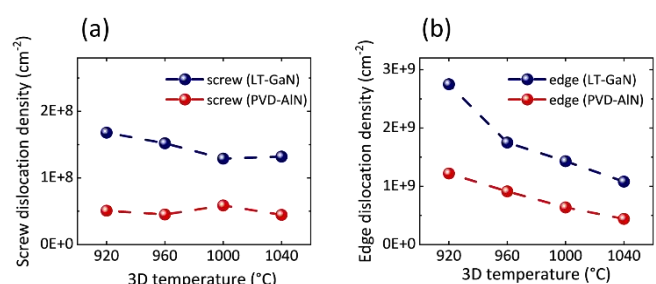

Fig. 7 Variation of (a) screw dislocation densities and (b) edge dislocation densities in GaN films grown on NPSS with LT-GaN and PVD-AlN NLs at different 3D temperatures.

Dark-field TEM was employed to analyze the dislocation distribution at the GaN grain coalescence boundaries. Fig. 8a-c and 7d-f show the dark-field TEM images of GaN films grown on NPSS using LT-GaN and PVD-AlN NLs, respectively. In Fig. 8a, it can be clearly seen that the LT-GaN NLs transform into small GaN grains after annealing at 3D temperatures. By contrast, the PVD-AlN NLs uniformly cover the NPSS, as shown in Fig. 8d. The superior thermal stability of PVD-AlN contributes to maintaining the orientation consistency of GaN grains, as supported by the XRD results in Fig. 5. Fig. 8b-c reveal that the dislocations in the GaN films grown on NPSS with LT-GaN NLs primarily originate from the coalescence region at the top of the cone-shaped nanopatterns. In this experiment, the cone-shaped nanopattern density of NPSS reaches approximately 4.5*10$^8$ cm$^{-2}$. Therefore, grain coalescence at the top region of the cone-shaped nanopatterns introduces a high density of dislocations in the GaN film grown on LT-GaN NLs. However, for GaN films grown on NPSS with PVD-AlN NLs, much fewer dislocation is observed at the top region of the nanopatterns, as seen in Fig. 8e-f. Combined with the XRD results in Fig. 5, the PVD-AlN NLs drastically reduce the tilt and twist angles of GaN grains compared to LT-GaN NLs. Accordingly, the coalescence process between small-angle tilted and twisted GaN grains grown on PVD-AlN NLs is likely to generate fewer dislocations.

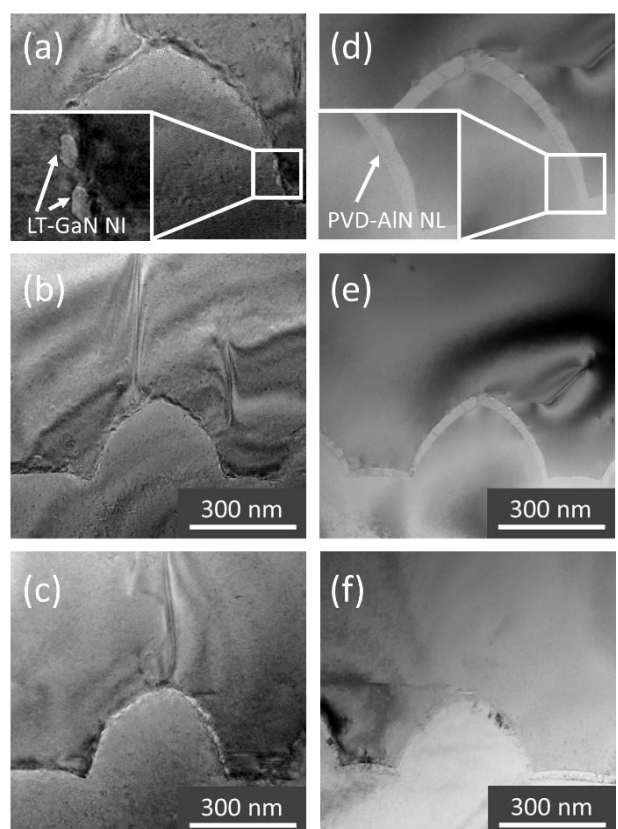

Fig. 8 The cross-sectional TEM images of GaN films grown on NPSS (a-c) with LT-GaN NLs and (d-f) with PVD-AlN NLs.

To explore the effect of grain coalescence on the stresses in GaN films, Raman spectroscopy was applied to measure the residual stress of samples A1-A4 and samples B1-B4. In fact, the observed residual stress in GaN films is the sum of the extrinsic compressive stress due to the thermal expansion mismatch to the sapphire and the intrinsic tensile stress which develops during grain coalescence[20]. Fig. 9a-b illustrate the Raman spectra of GaN films grown on NPSS with LT-GaN and PVD-AlN NLs at different 3D temperatures. The stress in GaN films can be calculated from the shift of the peak position of GaN $E_2(H)$ phonons. Specifically, the relationship between the in-plane stress $\sigma_{xx}$ and $E_2(H)$ phonon frequency shift $\Delta\omega$ in the GaN film can be expressed as: $\Delta\omega = K\sigma_{xx}$, where $K$ is the stress coefficient typically taking a value of 2.56 cm$^{-1}$/GPa on the sapphire substrates[40].

Fig. 9c illustrates the residual compressive stress of GaN films calculated from the $E_2(H)$ phonon frequency shift. As the 3D temperature increases from 920 to 1040 °C, the residual compressive stress of GaN on NPSS with LT-GaN NLs increases from 0.55 GPa to 0.98 GPa. This observation indicates higher 3D temperatures lead to a reduction in the tensile stress introduced during the coalescence process. Past studies indicate that the tensile stress induced by GaN grain coalescence during growth is strongly correlated with the size of the GaN grains[19, 20]. Specifically, the coalescence of smaller grains generates larger tensile stresses[19, 20]. Fig. 2-3 demonstrate that the size of the GaN grains increases as the 3D temperature rises. Therefore, the increase in residual compressive stress caused by the higher 3D temperatures can be mainly attributed to the decline in tensile stress resulting from the coalescence of larger grains. Similarly, the residual compressive stress in GaN on NPSS with PVD-AlN NLs increases from 1.09 GPa to 1.41 GPa. The residual compressive stress of GaN films on NPSS with PVD-AlN NLs is, on average, 0.5 GPa higher compared to that of GaN films on NPSS with LT-GaN NLs. The higher residual compressive stress is primarily attributed to the smaller in-plane lattice constants of PVD-AlN compared to GaN.

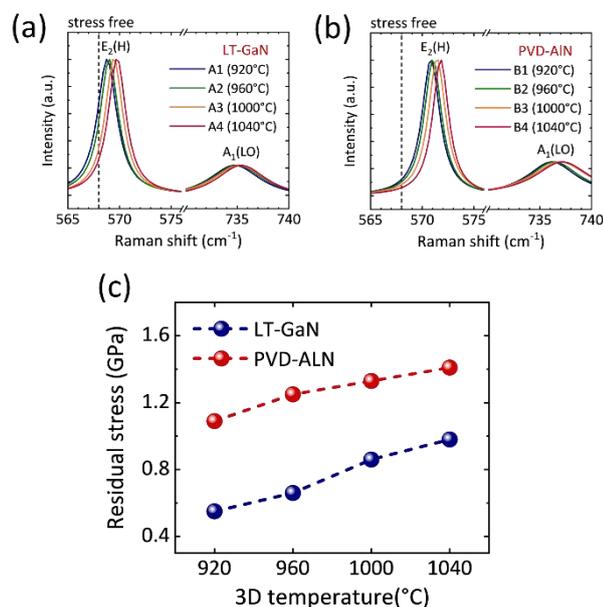

Fig. 9 Raman spectra of GaN films grown on NPSS with (a) LT-GaN NLs and (b) PVD-AlN NLs at different 3D temperatures. (c) The residual compressive stress calculated from the GaN $E_2(H)$ phonon peak position.

Raman mapping with a spatial resolution of 300 nm was employed to study the stress distribution of GaN films grown on NPSS with PVD-AlN NLs. Fig. 10a-b show the peak intensity map of GaN $E_2(H)$ phonon in the plane perpendicular and parallel to the c-axis direction of sample B4. Fig. 10c-d display the corresponding stress distribution obtained from the GaN $E_2(H)$ phonon peak position fitted by the Voigt function. In Fig. 10a, a clear NPSS periodic nanopattern can be observed in the peak intensity map of GaN $E_2(H)$ phonon in the plane perpendicular to the c-axis direction, which is mainly attributed to the difference in light extraction efficiency. Fig. 10c displays the corresponding stress distribution, indicating discrete dark spots and contiguous bright regions. The dark spots represent the locations where GaN grain coalescence occurs. As a result of the coalescence process, the residual compressive stress at these locations is reduced due to the introduction of tensile stresses. Specifically, the GaN $E_2(H)$ peak positions at the dark spots and bright regions diverge by about 0.5 cm$^{-1}$, as shown in Fig. 10e. These results indicate that the GaN at the dark spots locally introduces a tensile stress of about 0.2 GPa due to the grain coalescence process.

From the Raman mapping in the plane parallel to the c-axis shown in Fig. 10b and 10d, it can be seen that the stress distribution of the GaN near the substrate is relatively inhomogeneous. Fig. 10f illustrates the Raman spectra of selected points located in the dark and bright regions near the substrate. The GaN $E_2(H)$ peak positions at the dark and bright regions differ by about 0.5 cm$^{-1}$, corresponding to about 0.2 GPa, similar to that observed in the plane perpendicular to the c-axis. This stress inhomogeneity is mainly attributed to the tensile stresses induced by grain coalescence during the early stages of growth. After the completion of grain coalescence, the stress distribution in the 2D GaN layer becomes uniform. These Raman mapping results for the stress distribution in GaN films on NPSS align with our prior findings obtained through cathodoluminescence (CL) mapping[9].

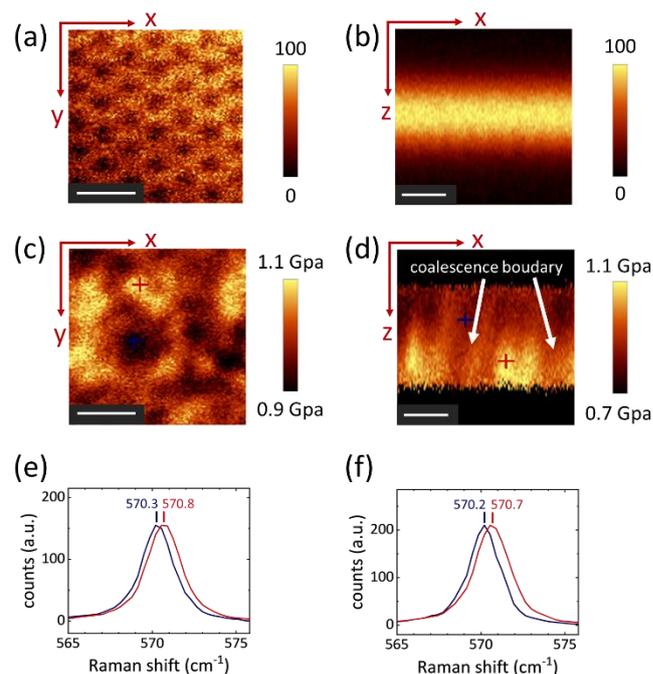

Fig. 10 The Raman mapping of GaN $E_2(H)$ phonon peak intensity in the plane (a) perpendicular and (b) parallel to the GaN c-axis direction of sample B4. The corresponding residual compressive stress distribution obtained from the GaN $E_2(H)$ phonon peak position fitted by Voigt function in the plane (c) perpendicular and (d) parallel to the c-axis direction of sample B4. (e-f) Raman spectra of the points selected in (c-d). The x-y plane is perpendicular to the c-axis direction, and the z-axis is parallel to the c-axis direction. All scale bars denote 1 μm.

Fig. 11a-b show schematic diagrams illustrating the effect of grain coalescence on the dislocation and stress evolution of GaN films grown on NPSS with LT-GaN NLs and PVD-AlN NLs. The LT-GaN NLs transform into small GaN grains after annealing at 3D temperatures, while the PVD-AlN NLs remain uniformly coated on the NPSS due to their high Al-N bonding energy. Concerning the LT-GaN NLs, a lower 3D temperature leads to a more uniform growth of 3D GaN grains, accompanied by a decrease in grain size and an increase in grain density. The coalescence of high-density, small-size GaN grains results in a substantial increase in edge dislocation density. Further, the coalescence of GaN grains introduces a greater tensile stress, thereby reducing the residual compressive stress.

In comparison to the LT-GaN NLs, the tilting and twisting of GaN grains grown on the PVD-AlN NLs are effectively suppressed. The coalescence process between small-angle tilted and twisted GaN grains grown on PVD-AlN NLs generates fewer dislocations. Notably, a significant decrease in dislocation can be observed at the coalescence boundary at the top of the cone-shaped pattern. In addition, the PVD-ALN layer introduces extrinsic compressive stresses due to the smaller in-plane lattice constants of AlN compared to GaN.

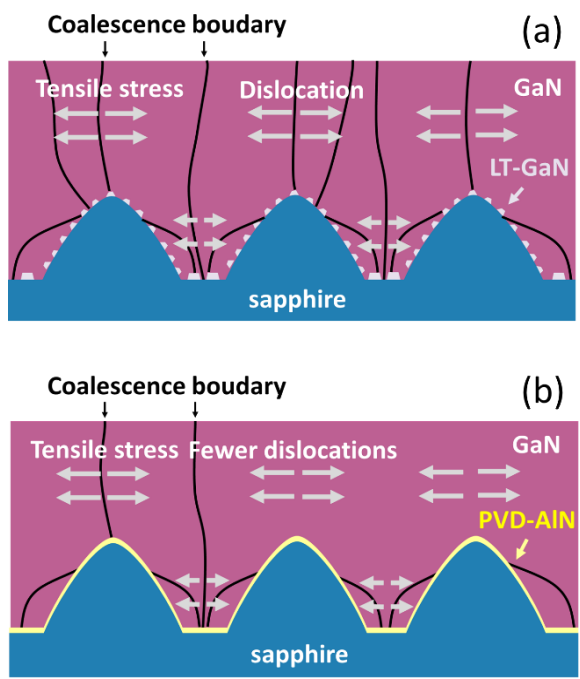

Fig. 11 Schematic diagram of dislocation and stress evolution in GaN films grown on NPSS with (a) LT-GaN and (b) PVD-AlN NLs.

## Conclusions

In summary, the effect of grain coalescence on the dislocation and stress evolution of GaN films grown on NPSS with LT-GaN NLs and PVD-AlN NLs is comparably investigated. The results show that higher 3D temperatures can reduce the edge dislocation density induced by grain coalescence in GaN films on NPSS. Meanwhile, higher 3D temperatures lead to a decrease in the tensile stress introduced by grain coalescence, consequently increasing the residual compressive stress in GaN films. Specifically for the LT-GaN NLs, higher 3D temperatures lead to severe Ostwald ripening of the 3D GaN grains, resulting in inhomogeneous growth. The PVD-AlN NLs significantly enhance the uniformity of the GaN grain growth and reduce the coalescence thickness. Furthermore, the PVD-AlN NLs effectively minimize the tilt and twist angles of the 3D GaN grains, thereby reducing the density of screw and edge dislocations at the grain coalescence boundary. In addition, the PVD-AlN NLs introduce additional residual compressive stress in the GaN layer due to the smaller lattice constant of AlN.

## Conflicts of interest

There are no conflicts to declare.

## Acknowledgements

This work was supported by.

## Notes and references